\documentclass{aastex}

\usepackage[percent]{overpic}
\usepackage{xcolor}
\usepackage{pict2e}
\usepackage{pbox}
\usepackage{amsmath,amssymb}
\usepackage{booktabs}
\usepackage{multirow}

\newcommand{\EQ}{\begin{equation}}
\newcommand{\EN}{\end{equation}}
\newcommand{\EQA}{\begin{eqnarray}}
\newcommand{\ENA}{\end{eqnarray}}

\newcommand{\JJ}{\mbox{\boldmath $J$} {}}
\newcommand{\UU}{\mbox{\boldmath $U$} {}}

\newcommand{\De}      {\mathrm{D}}
\renewcommand{\vec}[1]{\mbox{\boldmath{$#1$}}}

\newcommand{\Div}     {\vec{\nabla}\cdot}
\newcommand{\nab}     {\nabla}

\newcommand{\Strain}{\mbox{\boldmath ${\sf S}$} {}}

\newcommand{\Av}      {\vec{A}}
\newcommand{\Bv}      {\vec{B}}

\newcommand{\cs}      {c_{\rm s}}

\begin{document}

\title{Applying the weighted horizontal magnetic gradient method to a simulated flaring Active Region}
\author{M. B. Kors\'os\altaffilmark{1,2,3}, Piyali Chatterjee\altaffilmark{4} and R. Erd\'elyi\altaffilmark{1,3} }

\altaffiltext{1}{Solar Physics \& Space Plasma Research Center (SP2RC), School of Mathematics and Statistics, University of Sheffield, Hounsfield Road, S3 7RH, UK}

\altaffiltext{2}{Debrecen Heliophysical Observatory (DHO), Research Centre for Astronomy and Earth Sciences, Hungarian Academy of Science, 4010 Debrecen, P.O. Box 30, Hungary}

\altaffiltext{3}{Department of Astronomy, E\"otv\"os L\'or\'and University, P\'azm\'any P\'eter s\'et\'any 1/A, Budapest, H-1117, Hungary}

\altaffiltext{4}{Indian Institute of Astrophysics, II Block Koramangala, Bengaluru-560034, India}

\email{korsos.marianna@csfk.mta.hu, piyali.chatterjee@iiap.res.in, robertus@sheffield.ac.uk}

\begin{abstract}

Here, we test the weighted horizontal magnetic gradient ($WG_{M}$) as a flare precursor, introduced by \cite{Korsos2015}, by applying it to a magneto-hydrodynamic (MHD) simulation of solar-like flares \citep{Piyali2016}. The pre-flare evolution of the $WG_{M}$ and the behavior of the distance parameter between the area-weighted barycenters of opposite polarity sunspots at various heights is investigated in the simulated $\delta$-type sunspot. Four flares emanated from this sunspot. We found the optimum heights above the photosphere where the flare precursors of the $WG_M$ method are identifiable prior to each flare.
These optimum heights agree reasonably well with the heights of the occurrence of flares identified from the analysis of their thermal and Ohmic heating signatures in the simulation. We also estimated the expected time of the flare onsets from the duration of the approaching-receding motion of the barycenters of opposite polarities before each single flare. The estimated onset time and the actual time of occurrence of each flare are in good agreement at the corresponding optimum heights. This numerical experiment further supports the use of flare precursors based on the $WG_M$ method. 

\end{abstract}


\section{Introduction}

Solar active regions (ARs) are among the most investigated dynamic features on the Sun which are identified as a collection of strong positive and negative magnetic polarity elements (sunspots) in magnetograms. A sunspot is classified as $\delta$-type when the opposite magnetic polarities share a common penumbra \citep{Kunzel1960}. The magnetically complicated and highly dynamic $\delta$-type sunspot groups are more likely to produce flares and CMEs than the bipolar ones \citep{Tanaka1975, Sammis2000}. Studying the $\delta$-type sunspot groups may be a key element to reveal the characteristic temporal variations of the evolution of magnetic field prior to flares \citep{Leka1996, Lidia1997,Takizawa2015}. 

In the literature, several studies have addressed the spatial and temporal evolution of the flare-triggering phenomena. Observational and numerical investigations report that newly emerged magnetic flux \citep{Archontis2008, MacTaggart2009}, flux cancellation \citep{Livi1989, Wang1993, Sterling_etal2010, Green2011,Burtseva2013,Savcheva2012}, strong magnetic shear and the rotation \citep{Evershed1910,Kempf1910,Manchester2000, Manchester2004,Yan2008,DeVore2008, Selwa2012} along the polarity inversion line (PIL), the length or strong horizontal gradient across the PIL \citep{Schrijver2007,Falcon2008} seem all to be, with various degrees, candidates for flare and CME triggers. Also, a range of flare and CME models exist where the complex configuration of an AR are proposed to lead to solar eruptions \citep[][and references therein]{Aschwanden2005, Li2005,Shibata2011}.

Here, we investigate the evolution of the opposite magnetic polarities in a 3D numerical model of a $\delta$-type sunspot. We test the concept of the weighted horizontal magnetic gradient ($WG_{M}$) method proposed by  \cite{Korsos2015}, K15 hereafter, by analyzing the emerging magnetic flux which generates a series of flares in the simulation, first reported in \cite{Piyali2016}. In Section \ref{Pencil}, we outline briefly the simulation setup and the numerical code used. In section \ref{thermal_analysis}, we perform a detailed analysis of the flaring regions of the simulation in terms of Ohmic heating and temperature increase for comparison of the findings with the $WG_M$ method given in Section \ref{Analyses}. We describe the $WG_{M}$ method itself and present our analysis of the simulated AR, followed by summarising our findings.
Finally, we discuss on our results and draw conclusions in Sections \ref{conclusion}.

\section{The MHD model} \label{Pencil}

\begin{figure*}\centering
\begin{overpic}[width=0.315\textwidth]{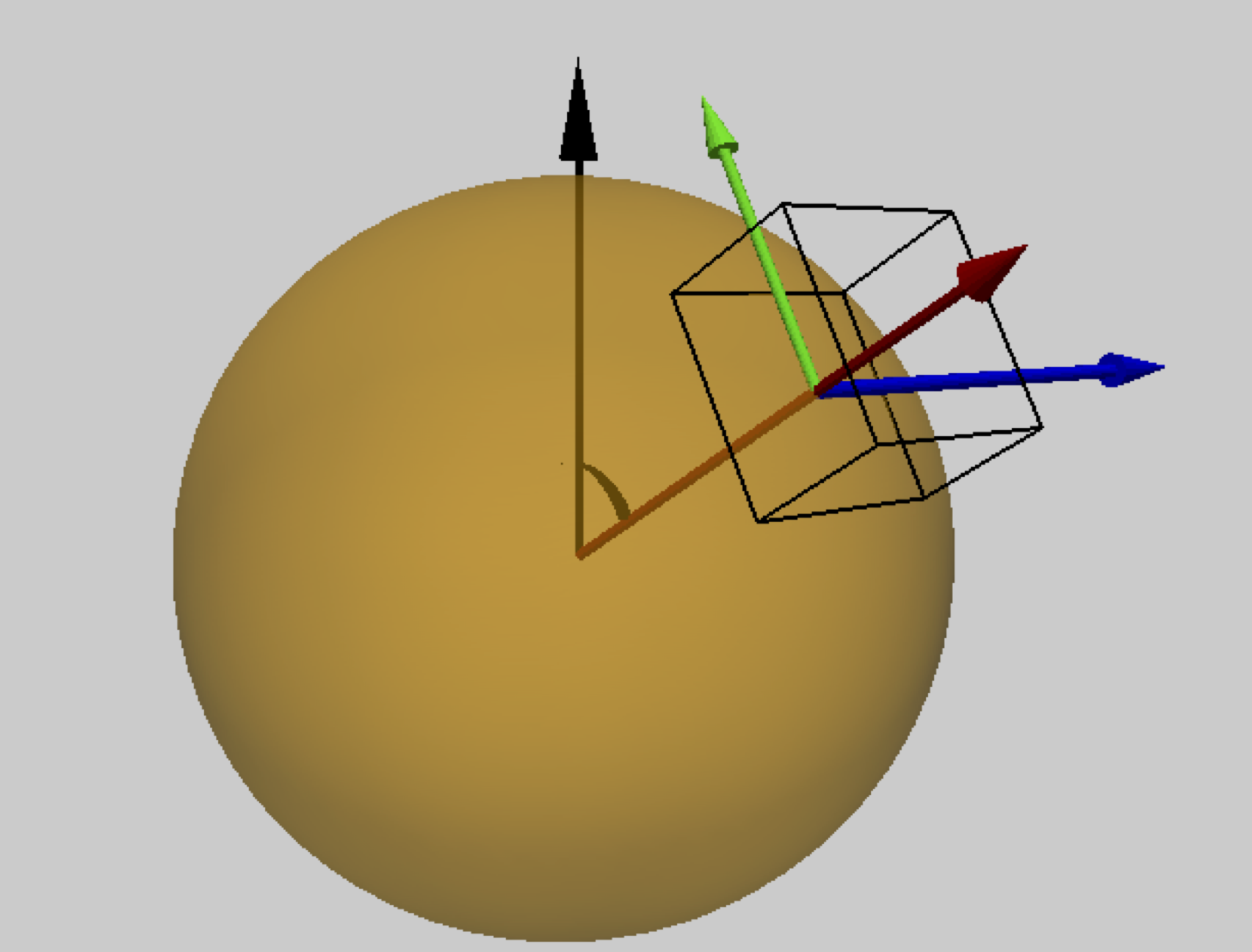}
\put(2,69){(a) }
\put(42,55){\Large{$\curvearrowleft$}}
\put(48,59){{$\Omega$}}
\put(50,40){{$\theta$}}
\put(90,49){{$\mathrm{\hat{x}}$}}
\put(59,70){{$\mathrm{\hat{y}}$}}
\put(83,58){{$\mathrm{\hat{z}}$}}
\end{overpic}
\begin{overpic}[width=0.45\textwidth]{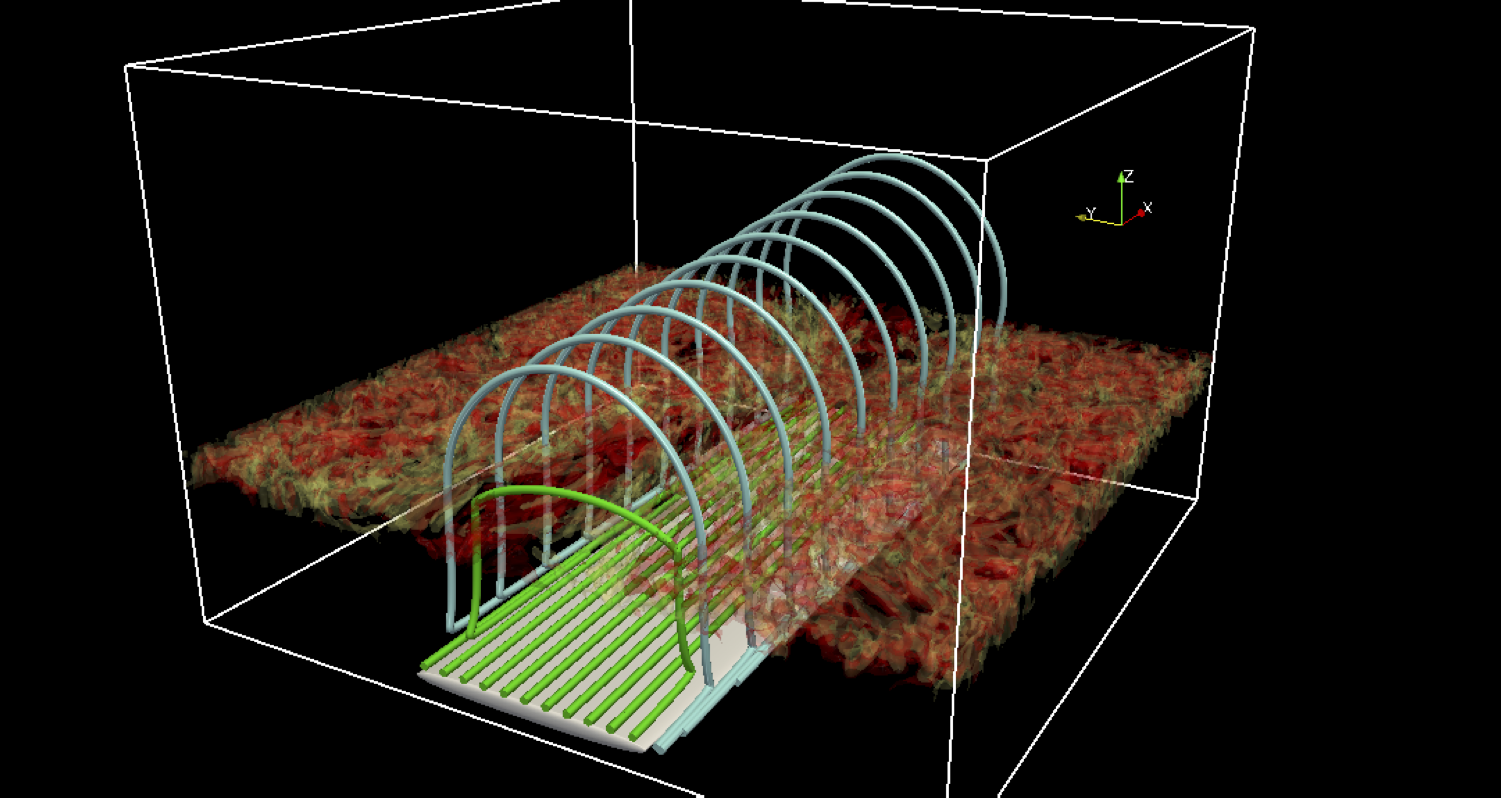}
\put(2,49){\color{white} (b) $t=0$}
\end{overpic}
\caption{\label{fig:cartoon} (a) The Cartesian simulation domain with respect to spherical coordinates. For visual clarity, the ratio of the horizontal extent of the box to the radius of the sphere in the picture is $10$ times larger than that used in the simulation. (b)  The initial state inside the box with a thin magnetic layer represented by the isosurface of $B\rho^{-1/4}$ (white). Few field lines in this layer are shown in green. Additionally, the ambient (arcade shaped) magnetic field lines are shown in cyan. The location of the photosphere is marked by convective granules represented by isosurfaces of $v_z$, with red (yellow) representing upward (downward) $v_z$. }
\end{figure*}

Our analysis is centered on the numerical case study reported in \cite{Piyali2016}. For completeness, we briefly describe the salient points of the model setup here. The computational domain consists of a box, with horizontal extents of -18 $  \mathrm{Mm} < \mathrm{{\it x}}, \mathrm{{\it y}} < $18  $\mathrm{Mm}$, and vertical one of -8.5 $\mathrm{ Mm}< \mathrm{{\it z}} <$ 16.5 $\mathrm{ Mm}$, rotating with a solar-like angular velocity $\Omega = 2.59\times10^{-6}$ s$^{-1}$, making an angle of $30^o$ with the vertical $z$-direction. A constant gravity, $g_z$, points in the negative $z$-direction.
The box is resolved using a uniformly spaced grid with d$\textrm{x}$ = d$\textrm{y}$ = 96 km and d$\textrm{z}=48$ km. The box may be thought to be placed at a colatitude $\theta$ on the surface of a sphere with its unit vectors, $\mathbf{\hat{x}}$, $\mathbf{\hat{y}}$, $\mathbf{\hat{z}}$ pointing along the local $\phi$, $-\theta$ and $r$ directions, respectively, as shown in Fig.~\ref{fig:cartoon}a. We use the fully compressible higher-order 
finite difference tool, the {\sc Pencil Code}\footnote{{https://github.com/pencil-code/}} for these calculations. This code is highly modular and can easily be adapted to different types of computational MHD problems.

The induction equation is solved for the magnetic vector potential, $\Av$, using the uncurled induction equation,  
\begin{equation}
  \frac{\partial\Av}{\partial t}
  = \UU\times\Bv - \eta\JJ +\nabla\Psi,
\end{equation}

where $\nabla\times\bf{A} = \bf{B}$ and $\eta$ denotes molecular magnetic diffusivity. Gauge freedom allows us to set $\Psi=0$ (Weyl gauge) at all times. The initial expression for the components of $\bf A$, corresponding to a horizontal magnetic sheet at $\textrm{z}_0=-7.75 ~\textrm{Mm}$ (shown by the white iso-surface in Fig. \ref{fig:cartoon}b) with the magnetic field vector, $\bf{B}$, strongly oriented in the {\it x}-direction, are given by,
 
$$A_\textrm{x}=q \varpi \Phi; A_\textrm{y}= -(\textrm{z}-\textrm{z}_0)\Phi; A_\textrm{z}= y \Phi,$$ where, $\Phi=B_0 R^2 \left[1-\textrm{exp}\left\{-\varpi^2/R^2\right\}\right]/\varpi$ with  $B_0 = 50 $ kG, $\varpi^2 = (a \mathrm{y})^2 + (\mathrm{z}-\mathrm{z}_0)^2$ and $a=0.1$. The horizontal extent of the sheet is about $-3~\mathrm{Mm}$ $<$ $y$ $<$ $3~\mathrm{Mm}$ and the maximum half-width, $R$, is 0.3 Mm at ${\bf y}=0$. With this value of $R$, the twist parameter, $q$, thus is 0.1 corresponding to an initially weak negative twist. We also introduce an ambient magnetic field in the form of a potential field arcade at $z$ $>$ 0, also shown in Fig.~\ref{fig:cartoon} {b}. The lower boundary at $z$=-8.5 Mm is closed and the top boundary at $z$= 16.5 Mm is open. The $x$-boundaries are periodic whereas the $y$-boundaries are perfectly conducting walls. Finally, we have for the entropy equation with temperature $T$, height-dependent thermal conductivity $K$, and turbulent diffusion, $\chi_t$,

\begin{eqnarray}
\label{eq:entropy}
\nonumber
   \rho T\frac{\De s}{\De t}
   =  \Div(K\nab T) + \Div(\rho T \chi_t \nab s)
      + \eta\mu_0\JJ^2\\
      + 2\rho\nu\Strain^2 
      -\rho^2\Lambda(T) + Q_{\textrm Cor} \,,
\end{eqnarray}

where the temperature is related to the sound speed by $\cs^2 = (c_p-c_v)\gamma T$.
The last two terms in Eq.~(\ref{eq:entropy}) are the radiative cooling and coronal heating terms, respectively. 

We include explicit height-dependent viscosity in the velocity equation, $\nu/\nu_0=1+f(1+\tanh \left\{ (z-z_{1})/w \right\})$, whereas magnetic diffusivity, $\eta/\eta_0$, and isotropic thermal conductivity, $K/K_0$, vary as $(\rho_{\tiny \textrm{in}}/\rho_0)^{-1/2}$, with, $f=150$, $z_{1}= 2$ Mm, $w=1.5$ Mm, $\nu_0= 2\times10^{10}$ cm$^2$ s$^{-1}$, $\eta_0=10^4$ cm$^2$ s$^{-1}$, $K_0=5\times10^4$ cm$^2$ s$^{-1}$, 
and $\rho_{\tiny \textrm{in}}$ is the initial density. The turbulent diffusion, $\chi_t = 10^{11}$ cm$^2$ s$^{-1}$ for $z < 0$ and tends to zero above that. Additionally, we use hyper-dissipation and shock viscosity proportional to positive flow convergence, maximum over three zones, and smoothed to second order. A density diffusion of $10^{11}$ cm$^2$ s$^{-1}$ is also included throughout since the plasma-$\beta$ reaches values $\sim 10^{-3}$. After a time, $t = 220$ min in the simulation, we have increased the value of density diffusion to $10^{12}$ cm$^2$ s$^{-1}$ and $f=300$ to prevent the velocities from going to infinity in the code.

\section{Analysis of temperature and Ohmic heating in the simulation}\label{thermal_analysis}

The simulation was ran for 263 min of solar time starting from the initial state shown in Fig.~\ref{fig:cartoon}b. It takes about 145 min from the start for the initial magnetic sheet to break up, rise and emerge through the surface like a newly emerging active region. Afterwards, there were four eruptions identified as flares ($B_{1}$, $C_{1}$, $B_{2}$ and $C_{2}$) with magnetic energy released equal to $3.3\times10^{29}$ ergs, $1.7 \times 10^{30}$ ergs, $2\times 10^{29}$ ergs and $2.3\times10^{30}$ ergs at simulation onset times $t=167.5$ mins ($B_{1}$), $t=197.2$ min ($C_{1}$), $t=215.03$ mins ($B_{2}$), and $t=240.2$ mins ($C_{2}$), respectively. Comparing with the estimates made by \cite{Isobe_etal2005} for a C-class flare that occurred on November 16, 2000, we conclude that the first and the third  flares can be categorised as Geostationary Operational Environment Satellites (GOES) B-class, whereas the second and fourth as GOES C-class for the amount of x-ray flux emitted. {In Table~\ref{tab:flareonset}, we show the onset times, energy released and estimated reconnection height for each flare. The onset times of the flares are obtained from the temporal evolution of the magnetic energy. For example, Fig.~\ref{fig:B2energy} shows the evolution of magnetic energy (black) and the Poynting flux (red) in a sub-domain surrounding the $B_{2}$ flare. The flare onset for this flare occurs at t = 215.03 min when there is a local maximum in the energy curve and the slope of the energy curve starts to change sign from positive to negative, with the energy decreasing rapidly.
{In order to differentiate the flare onset signal from other fluctuations, we combine the information of change of slope of energy versus time with the first appearance of the flashes of high temperature}
 in the accompanying animations files at three different heights. Also we use the information from time of occurrence of the bipolar reconnection jets in Fig. 4(b) of Chatterjee et. al. (2016) for the $B_{1}$, $C_{1}$ and $C_{2}$ flares which matches with the time from the energy curves in Fig 4(a) of the same paper. The magnetic energy, $\delta \mathcal{E}_B$, released during the $B_2$ flare is calculated to be $2\times10^{29}$ ergs. The $\delta \mathcal{E}_B$ values for the $B_{1}$ flare and the $C_{1}$ flares were given in \cite{Piyali2016} as well as in Table.~\ref{tab:flareonset} for completeness. The Poynting flux into the area surrounding the flare also decreases rapidly after $t=215.03$ min and becomes close to zero.

In Fig.~\ref{fig:mag_temp}, we show the contours of temperature anomaly, as $\Delta T$, relative to the horizontal average, denoted $\overline{T}(z)$, at three different heights, $z=0.59$ Mm, $z=1.28$ Mm and $z=3.24$ Mm for all the flares we study in the simulation. A positive (negative) $\Delta T$  implies that the local temperature is greater (less) than $\overline{T} (z)$ of the horizontal layer. It is clear from the temperature indicator that the $B_{2}$ flare occurred much below $z=3.24$ Mm, whereas some signatures of the $B_{1}$ and $C_1$ flare can still be detected at this height. Moreover, the $B_{2}$, $C_1$ and $C_2$ flares can be detected much lower in the atmosphere, e.g. as low as at $z=0.59$ Mm, contrary to $B_1$ which does not show any brightening at this height at $t=168.89$ min. However, from a later time, $t=170.56$ min, we start seeing the flare brightening for $B_1$ at the height $z=0.59$ Mm (see the accompanying animation file \href{ftp://ftp.iiap.res.in/piyali.chatterjee/f3.mp4}{\color{blue}f3.mp4} after $t=167$ min in the online journal). This means that the reconnection for flare $B_1$ was actually initiated higher up and it took $\sim 2$ min for the reconnection current sheet to stretch downwards, thus increasing the temperature of the lower layers.  {Similarly, from the middle panel of the animation \href{ftp://ftp.iiap.res.in/piyali.chatterjee/f3.mp4}{\color{blue}f3.mp4}, after $t=167$ min, one can also spot the reconnection jet before the  appearance of the bright inverse-shaped flux rope. This may mean that reconnection for the $B_1$ flare was actually initiated somewhere between $0.59- 1.28$ Mm.} Also note that the $B_1$ and the $B_{2}$ flares erupted over different regions of the simulation domain. In general, all flares appear bright in terms of $\Delta T/\overline{T}(z)$ at $z=1.28$ Mm. The last flare, $C_2$, is most likely a filament eruption as evident from two neighbouring inverse-S shaped dark filamentary structures in the $\Delta T/\overline{T}(z)$ contour plot at all heights. The evolution and eruption of this filament-like structure is shown in Fig.~6 of \cite{Piyali2016}. There, one sees some smaller bright regions surrounding the dark filaments at the heights $z=0.59, 1.28$ Mm. A corresponding bright region at $z=3.28$ Mm is not so prominent likely because of a large coronal conductivity used in the MHD equations after $t=220$ mins. 

\begin{table}[h!]
  \centering
  \caption{\label{tab:flareonset} Summary of the energy released, onset times, durations and estimated occurrence heights for the four flares.}
  \label{tab:table1}
  \begin{tabular}{@{}lrrrr@{}}
    \toprule
  & {\bf $B_1$} & {\bf $C_1$} & {\bf $B_2$} & {\bf $C_2$}\\
    \midrule
 $\delta\mathcal{E}_B$ ($10^{29}$ ergs) & 3.3 & 17.0&2.0 &23.0\\
   Onset time (min)& {$167.5$} & $197.2$& $215.03$&$240.2$\\
    Duration (min) & $5.0$ & $25.0$ & $13.0$&$>23.0$\\
    {\multirow{2}{2.4 cm}{Height range (Mm):}} & & & & \\ 
    & & & &\\
$\Delta T/\bar{T}(z)$& $> 0.6$ & $<3.24$&$<1.28$ & $\leq 3.24$\\
 Peak of $Q_{FL}$ & $0.4-1.5$& $2.5$& $0.3-0.5$&$3.0$ \\
$WG_M$ & $0.3-0.4$ & $2.3-2.9$& $0.5$ & $1.2-1.8$\\
    \bottomrule
  \end{tabular}
\end{table}

\begin{figure}[h!]
\centering
\begin{overpic}[width=0.5\textwidth]{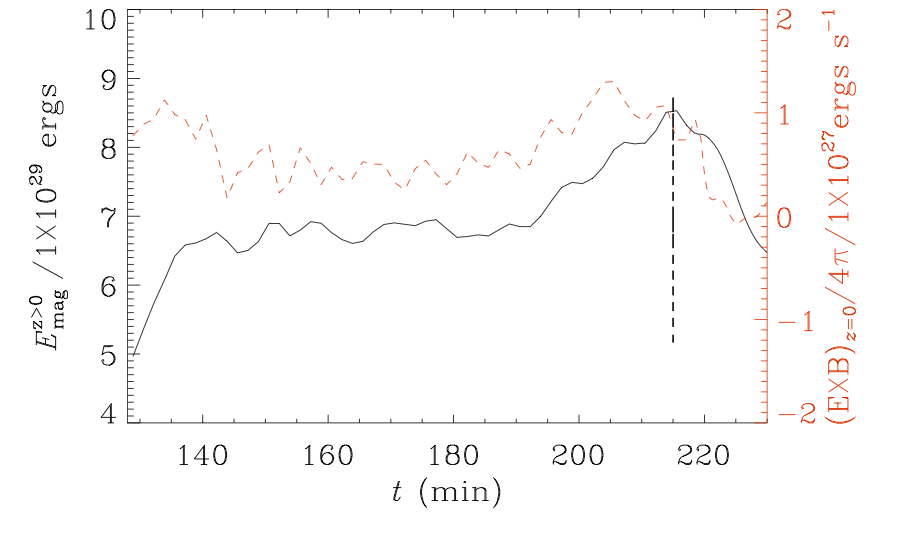}
\end{overpic}
\caption{\label{fig:B2energy} Evolution of magnetic energy (black line) and Poynting flux (red line) over an area surrounding the B$_2$ flare. The dashed vertical line denotes $t=215.03$ min, the onset time of the flare.}
\end{figure}

\begin{figure*} [ht!]
\centering
\begin{overpic}[width=0.22\textwidth]{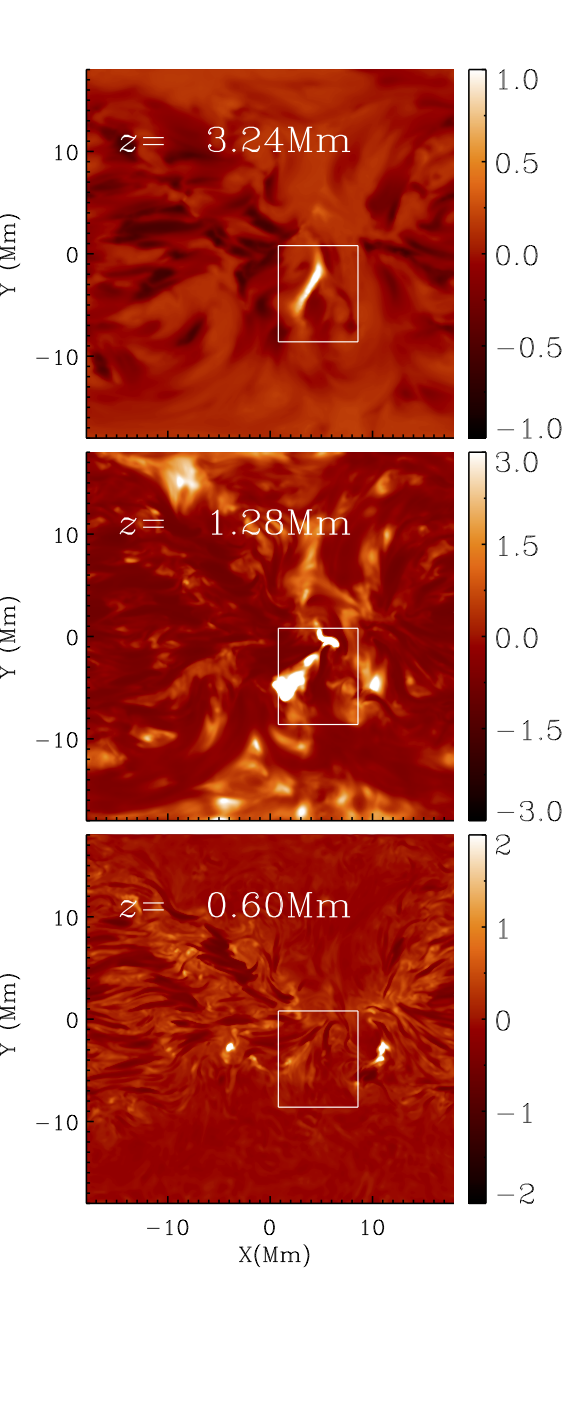}
\put(5,96){(a) $t=168.89$ min ($B_1$)}
\linethickness{2pt}
\put(15,75){\color{green}\vector(2.5,1.5){5}}
\put(15,50){\color{green}\vector(2.5,1.5){5}}
\put(29,58){\color{black}\vector(-2.5,-1.5){5}}
\end{overpic}
\begin{overpic}[width=0.22\textwidth]{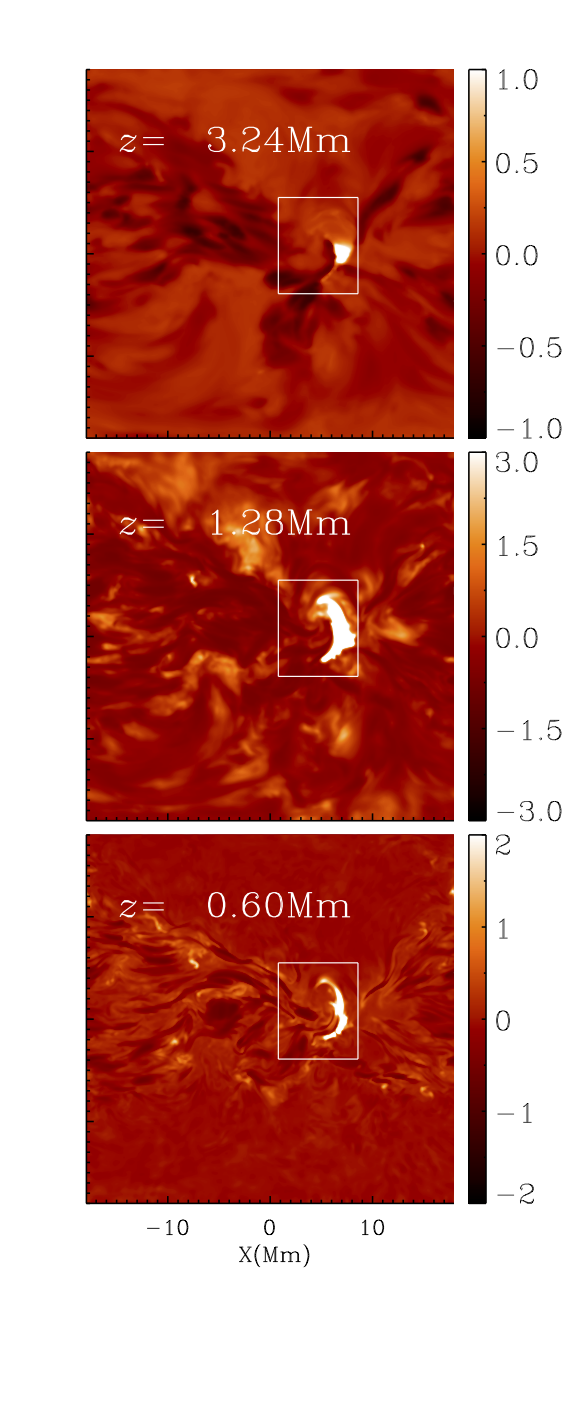}
\put(5,96){(b) $t=198.89$ min ($C_1$)}
\end{overpic}
\begin{overpic}[width=0.22\textwidth]{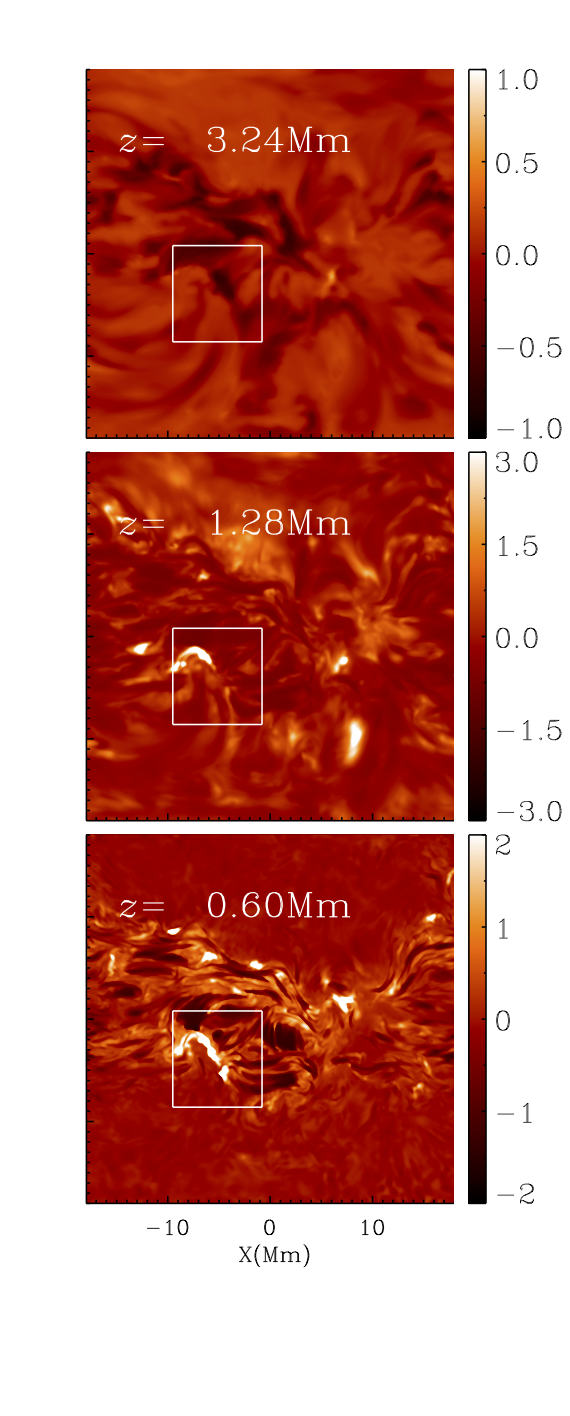}
\put(5,96){(c) $t=218.03$ min ($B_2$)}
\end{overpic}
\begin{overpic}[width=0.22\textwidth]{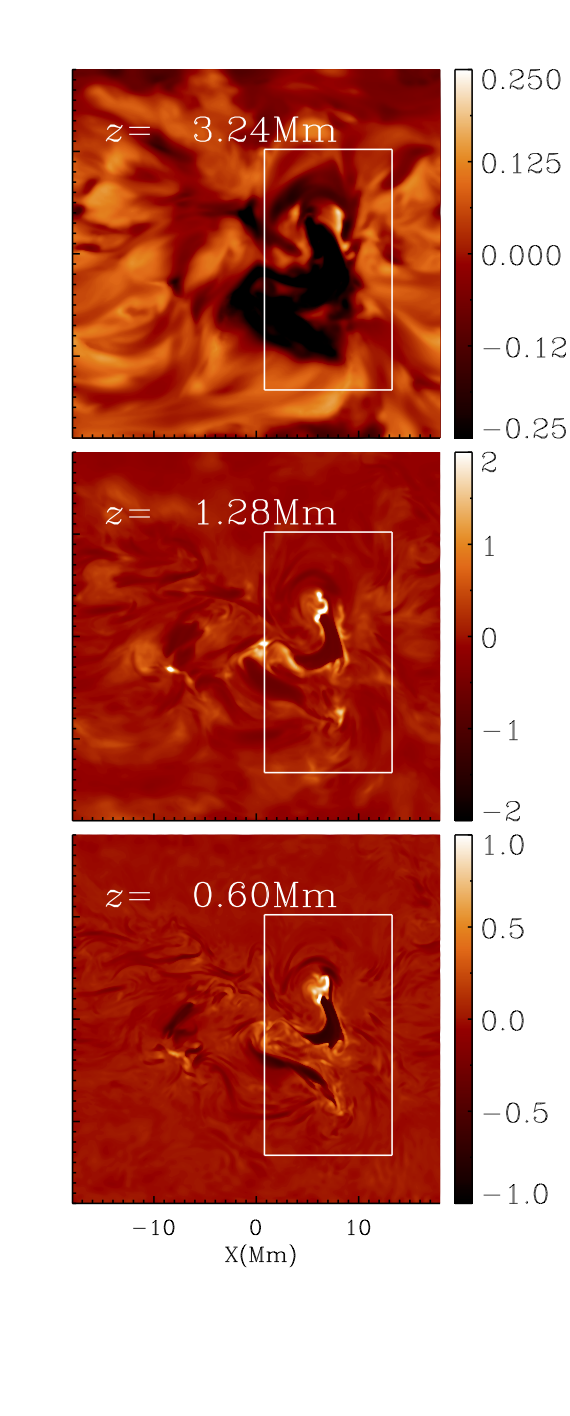}
\put(5,96){(d) $t=240.0$ min ($C_2$)}
\end{overpic}
\caption{\label{fig:mag_temp} (a) The ratio of the local temperature anomaly, $\Delta T$, to the horizontal average temperature, $\overline{T}(z)$, during the $B_1$ flare at three different heights indicated. A value of $\Delta T/\overline{T}(z) =s$ implies that the local temperature is $(s+1)\times \overline{T}(z)$. The green arrow (left column, top and middle panels) denotes the outward reconnection jet while the black arrow (left column, middle panel) denotes the hot channel of the magnetic flux rope. (b), (c), (d) are similar to (a) but for the $C_1$, $B_2$ and $C_2$  flares, respectively. White boxes demarcate the region surrounding the flares. This figure is available as an animation at this \href{ftp://ftp.iiap.res.in/piyali.chatterjee/f3.mp4}{\color{blue}link}. The animation starts at t=112.22 minutes and ends at 250.72 minutes. The 138.5 minutes of simulation time is compressed into a 45 sec animation.}
\end{figure*}

\begin{figure}[ht!]
\centering
\begin{overpic}[width=0.45\textwidth]{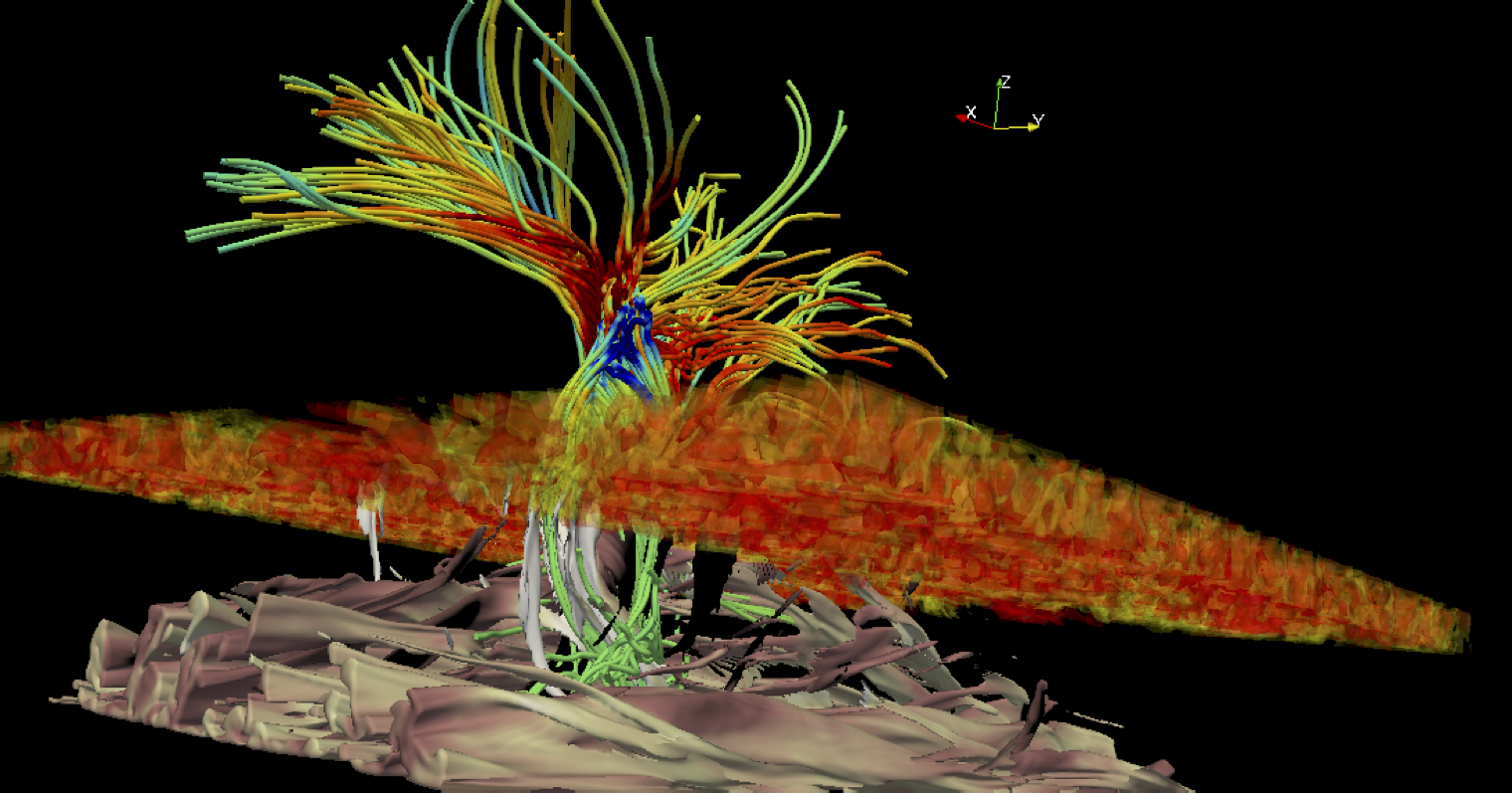}
\put(2,45){\color{white} $t=198.89$ min}
\end{overpic}
\caption{\label{fig:snap}
The simulation domain at the time of the C$_1$ flare. The field lines are colored according to plasma velocity orientation along them with red (blue) representing upward (downward) velocity.} 
 \end{figure}
\begin{figure*}[ht!]

\begin{overpic}[width=0.45\textwidth]{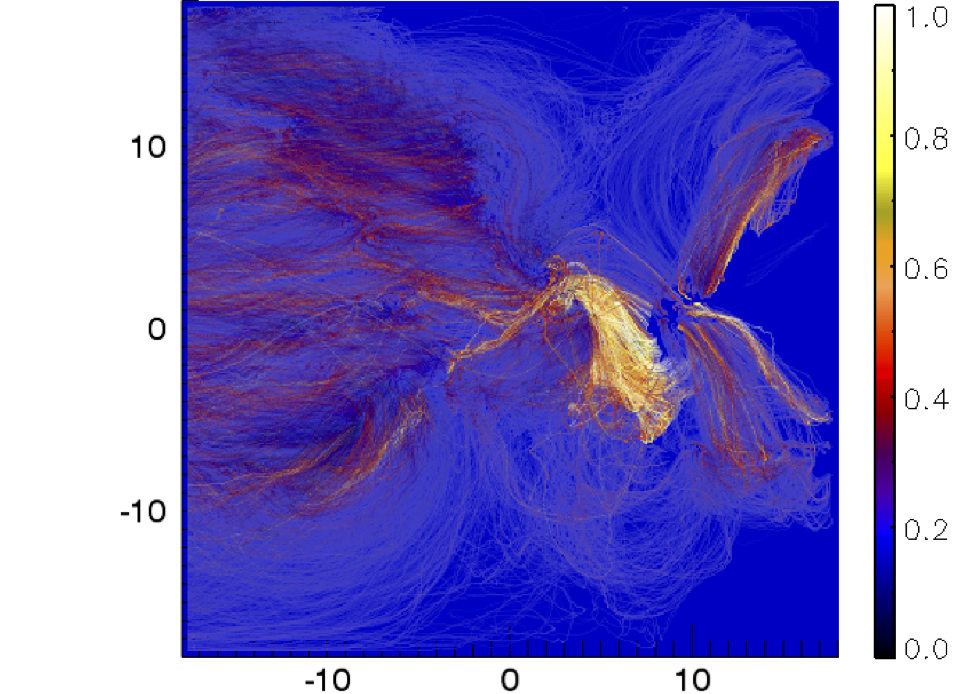}
\put(20,72){(a) $t = 167.22$ min}
\put(25,65){\color{white} $\sqrt{Q_{\mathrm{FL}}}$}
\put(10,25){\rotatebox{90}{Y (Mm)}}
\put(120,-4){
\includegraphics[width=0.48\textwidth]{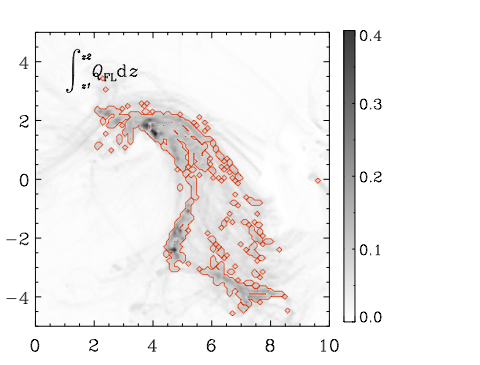}}
\end{overpic}\\
\\
\begin{overpic}[width=0.45\textwidth]{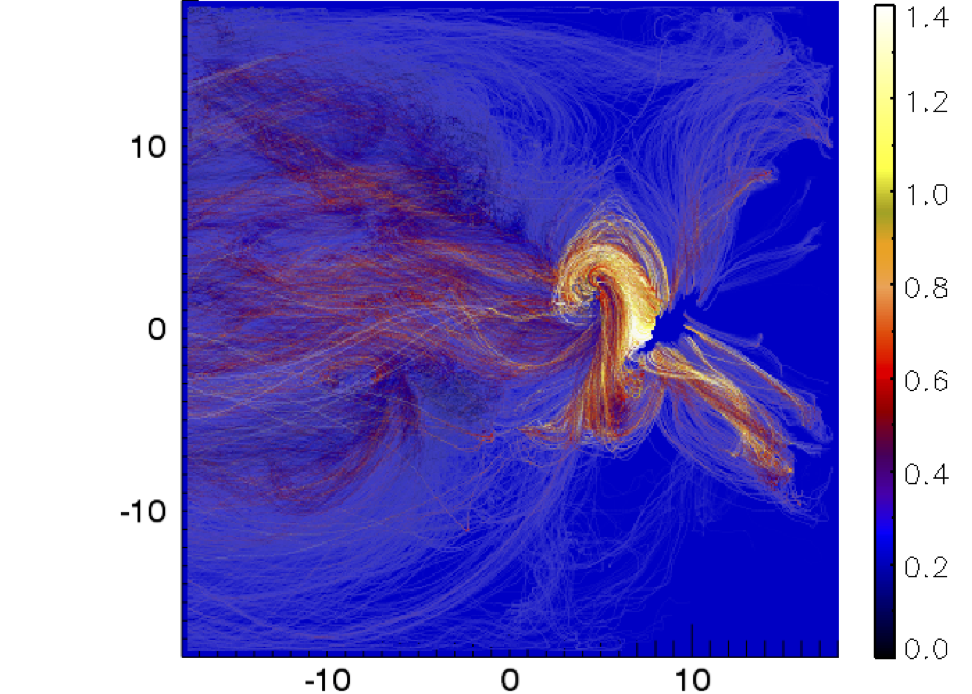}
\put(20,72){(b) $t = 197.22$ min}
\put(25,65){\color{white} $\sqrt{Q_{\mathrm{FL}}}$}
\put(10,25){\rotatebox{90}{Y (Mm)}}
\put(45,-5){{X (Mm)}}
\put(120,-8){
\includegraphics[width=0.48\textwidth]{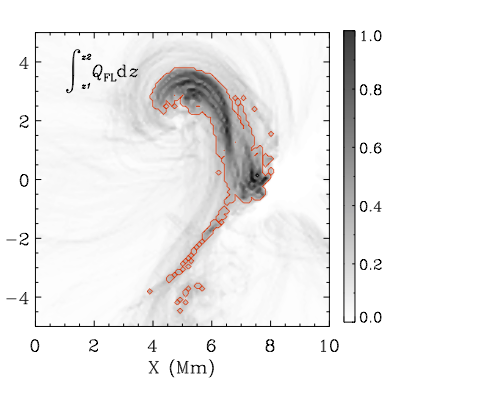}}
\end{overpic}
\\
\caption{\label{fig:qflB1C1}(a) {\em Left panel}: The square-root of the Ohmic field line heating, $Q_{\mathrm{FL}}$, in arbitrary units at $t=167.22$ min (onset of $B_1$ flare). {\em Right panel}: A zoomed-in view of $Q_{\mathrm{FL}}$ integrated between heights $z_1=616$ km and $z_2=11.6$ Mm. The red contour boundary denotes the region where $\int Q_{\mathrm{FL}} \mathrm{d}z > Q_c (=0.125)$ (see text). (b) Similar to (a) but just before onset of the $C_1$ flare and with $Q_c=0.5$. The left panels are available as an animation at this \href{ftp://ftp.iiap.res.in/piyali.chatterjee/f5.mp4}{\color{blue}link}. The animation starts at t=150.56 minutes and ends at 255.06 minutes. The 104.5 minutes of simulation time is compressed into a 46 sec animation.
}
\end{figure*}

\begin{figure*}[ht!]
\begin{overpic}[width=0.45\textwidth]{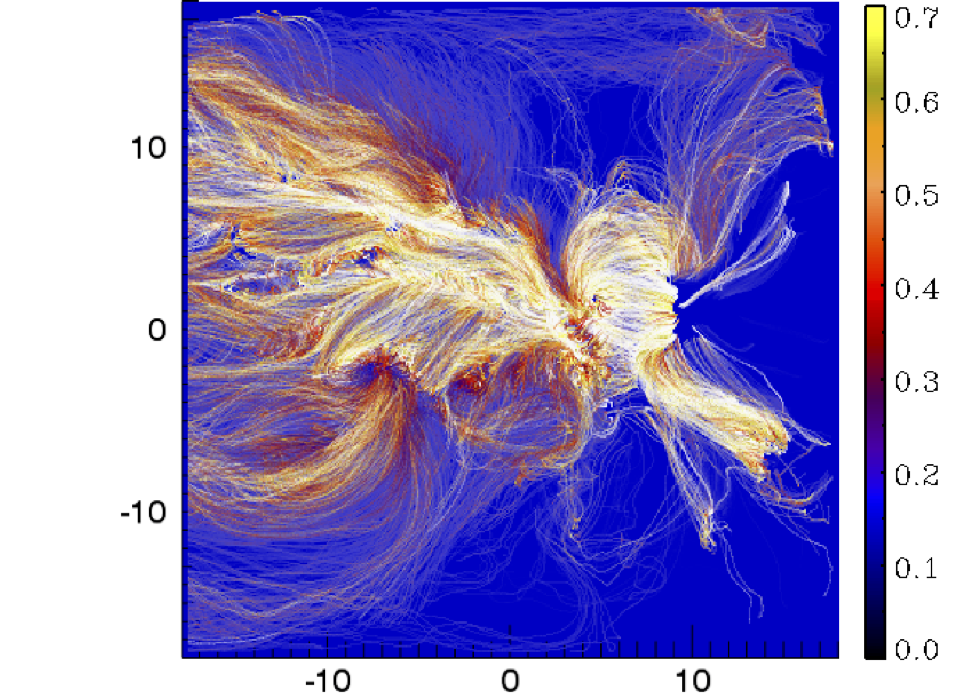}
\put(20,72){(a) $t = 217.2$ min}
\put(25,65){\color{white} $\sqrt{Q_{\mathrm{FL}}}$}
\put(10,25){\rotatebox{90}{Y (Mm)}}
\put(120,-4){
\includegraphics[width=0.48\textwidth]{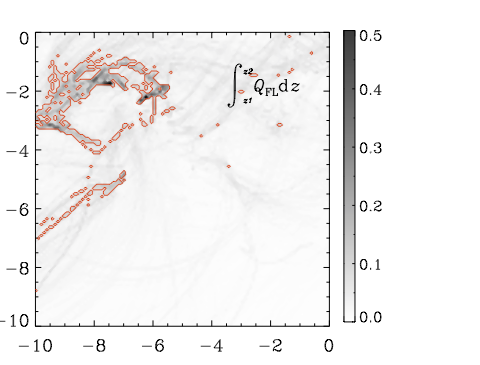}}
\end{overpic}\\
\\

\begin{overpic}[width=0.45\textwidth]{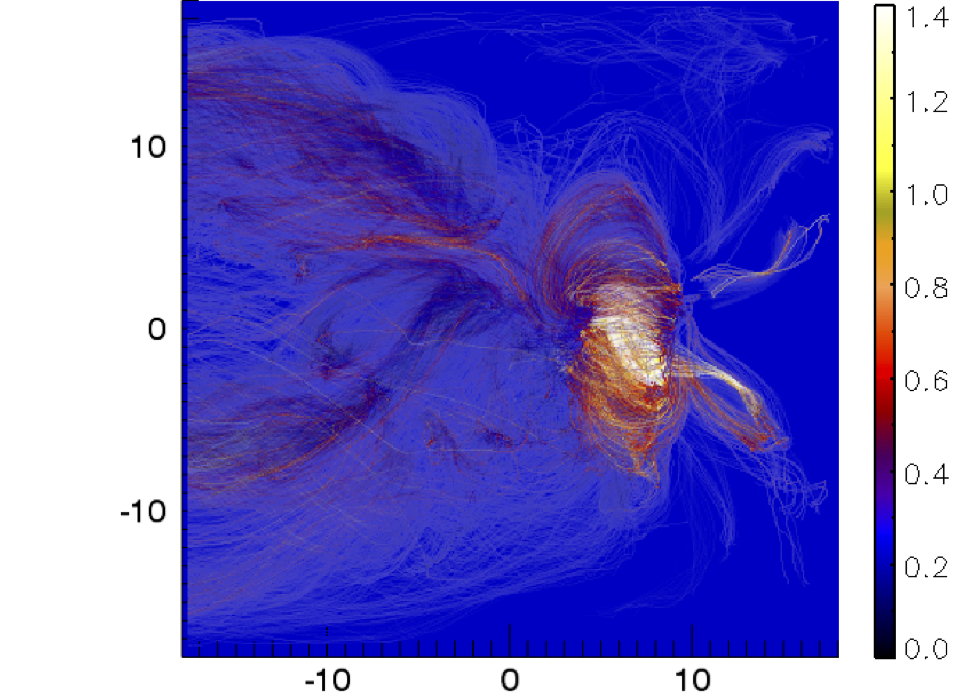}
\put(20,72){(b) $t = 239.06$ min}
\put(25,65){\color{white} $\sqrt{Q_{\mathrm{FL}}}$}
\put(120,-8){
\includegraphics[width=0.48\textwidth]{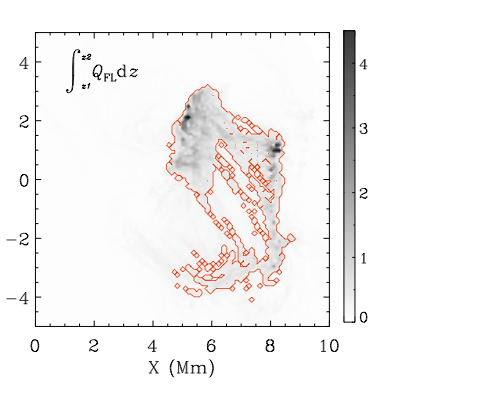}}
\put(10,25){\rotatebox{90}{Y (Mm)}}
\put(45,-5){X (Mm)}
\end{overpic}
\\
\caption{\label{fig:qflB2C2} (a) {\em Left panel}: The square-root of the Ohmic field line heating, $Q_{\mathrm{FL}}$, in arbitrary units at $t=217.2$ min ($B_2$ flare). {\em Right panel}: A zoomed-in view of $Q_{\mathrm{FL}}$ integrated between heights $z_1=616$ km and $z_2=11.6$ Mm. The red contour denotes the region where $\int Q_{\mathrm{FL}} \mathrm{d}z > Q_c (=0.125)$ (see text). (b) Similar to (a) but just before the onset of the $C_2$ flare, with $Q_c=0.5$. The left panels are shown in the Figure 5 animation.}
\end{figure*}

A simulation snapshot at $t=198.8$ min showing the reconnection jet along reconnecting field lines during the occurrence of the $C_1$ flare is shown in Fig.~\ref{fig:snap}. One important feature of this simulation seen in this figure is the self-consistent formation of helical and buoyant magnetic flux tubes under the action of magnetic  buoyancy instability on a thin magnetic sheet in the presence of rotation and stratification.  Furthermore, the flux tubes formed are non-uniformly twisted along their length and therefore can emerge out of the photosphere only at certain points where the twist is large. This alleviates the need to use uniformly twisted flux tubes with twist as a free initial parameter. 
We will describe our detailed analysis of flare initiation sites later in the text.

We have used ideal gas thermodynamics in this simulation without solving for detailed radiative transfer and without taking into account the effects of ionisation. Also, in order to keep the simulation stable at low plasma-$\beta$, we have used higher dissipation. All these approximations can make the temperature in the simulation a less reliable indicator. Alternatively, we can also estimate the Ohmic heating of field lines above the photospheric height in the simulation using a method similar to the one illustrated in \cite{Cheung_DeRosa2012}. The Ohmic heating term in Eq.~(\ref{eq:entropy}) is given by 
$\eta \mu_0 {\bf J}^2$. If, however, we were to write an equation for the temperature, $T$, instead of for entropy, $s$, the Ohmic heating term will be given by, $\eta\mu_0{\bf J}^2/\rho C_v$. Assuming that the thermal conductivity along magnetic field lines far exceeds the isotropic thermal conductivity in the solar corona we can assign a quantity, $\tau_{\mathcal{L}}$, to a line-tied field line $\mathcal{L}$ where, 
$$\tau_{\mathcal{F}}=\frac{\mu_0}{c_v\mathcal{L}}\int_{\mathcal{F}} 
\frac{\eta{\bf J}^2}{\rho} d\vec{l}.$$

Here, $c_v$ is the specific heat capacity at constant volume and $d\vec{l}$ is an infinitesimal distance along the field line $\mathcal{F}$ of length $\mathcal{L}$ between the line-tied ends at the photosphere. We trace about $10^5$ field lines through all the points on the photosphere where $B_z > 200$ $G$ and assign a unique $\tau_{\mathcal{F}}$ to all the field lines. If the field line crosses any of the side boundaries or the top boundary then we set $\tau_{\mathcal{F}}=0$ for that field line. Now, any magnetic field line will traverse through many grid cells in the computational domain. For, each grid cell we define the increment in the value of Ohmic heating denoted, $Q_{\mathrm{FL}}(x, y, z)$, by,

$$\mathrm{d}Q_{\mathrm{FL}} = \tau_{\mathcal{F}} \mathrm{d}x\mathrm{d}y.$$

Hence, the net heating due to field lines, $Q_{\mathrm{FL}}$, for any grid cell will be the sum of $\tau_{\mathcal{F}}$ for all field lines passing through that cell. A region like a current sheet or a flux rope will appear bright in $Q_{\mathrm{FL}}$ as all field lines passing through it carry large currents and so, have a large value of $\tau_{\mathcal{F}}$. The 3-dimensional Ohmic heating, $Q_{\mathrm{FL}}(x, y, z)$ at the onset time of all the four flares - $B_1, C_1, B_2,$ and $C_2$- as viewed from the $z$-direction are shown in the left panels of Figs.~\ref{fig:qflB1C1} and \ref{fig:qflB2C2}. The field lines carrying the largest currents appear brighter than the surroundings and can easily be spotted in all these panels. The regions surrounding these field lines are likely to be hot because of Ohmic dissipation. An animation file -- qfl.mp4 -- for the entire simulation duration is provided with the online journal. The epochs of the appearance of bright current carrying lines in the animation show excellent correlation with the flare onset times calculated using the magnetic energy release (as a function of time) in \cite{Piyali2016}. The right panels of Figs.~\ref{fig:qflB1C1} and \ref{fig:qflB2C2} depict the quantity, $Q_{\mathrm{FL}}$ integrated between heights $z_1=660$ km and $z_2=11.6$ Mm at the snapshot time indicated and can be compared to observational EUV or soft x-ray images of heated coronal loops. Once we have the $z$-integrated heating $Q_{\mathrm{FL}} (x,y)$ at each time, we now choose a set of points for each simulation snapshot on the $xy$-plane where $Q_{\mathrm{FL}} (x,y) > Q_c$, a critical value. This critical value has been chosen as 0.5 for flares $C_1$ and $C_2$ and 0.125 for $B_1$ and $B_2$ flares so as to obtain a mask of about 500 points for each case. The outer boundary of this masked region is denoted by a red contour in the right panels. In these panels, to show the heated regions clearly, we have zoomed into the region surrounding flares. 

In Fig.~\ref{fig:qflbar}, we show the heating function, $\overline{Q}_\mathrm{FL}(z)$ as a function of $z$, obtained by averaging $Q_\mathrm{FL}$ over all the points inside the boundary of the red contours shown in the right panels of Fig.~\ref{fig:qflB1C1} and ~\ref{fig:qflB2C2} for all the four flares.  Moreover, we have temporally averaged the $\overline{Q}_\mathrm{FL}$ curves for simulation snapshots between an interval $\pm 2.8$ mins around the onset time. Just before the onset of any flare when the function $\overline{Q}_\mathrm{FL}$ peaks at a certain height, we can conclude that the flare was likely initiated at that height. For flare $B_1$, the $\overline{Q}_\mathrm{FL}$ shows a plateau between $0.1-2$ Mm, whereas for flare $B_2$, we see a clear peak at 0.5 Mm. Flares $C_1$ and $C_2$ also have plateaus between $0.1-3$ and $0.1-4$ Mm, respectively. Also, the peaks (of $Q_\mathrm{FL}$)  for flares $C_1$ and $C_2$ appear at heights 2.6 and 3 Mm above the photosphere, respectively. From these results, and aided with the online animation at corresponding heights - \href{ftp://ftp.iiap.res.in/piyali.chatterjee/f3.mp4}{\color{blue}f3.mp4} - we can conclude that the flare $B_2$ was likely initiated at 0.5 Mm whereas flares $C_1$ at $\sim 2.6$ Mm and $C_2$ at $\sim 3$ Mm, respectively. For the flare $B_1$, because of the flat plateau without any pronounced peaks, we can only conclude that it was initiated below the height of $1.5$ Mm. The $WG_M$ method will be applied at different heights of the simulation, with the goal of understanding its behavior relative to the derived heights of the flare initiation using Ohmic heating as well as temperature signatures. This spatial information, gained from analysis of this simulation, will be compared with the output of the $WG_M$ analysis as function of height in the next section.

\begin{figure}[h!]
\begin{overpic}[width=0.45\textwidth]{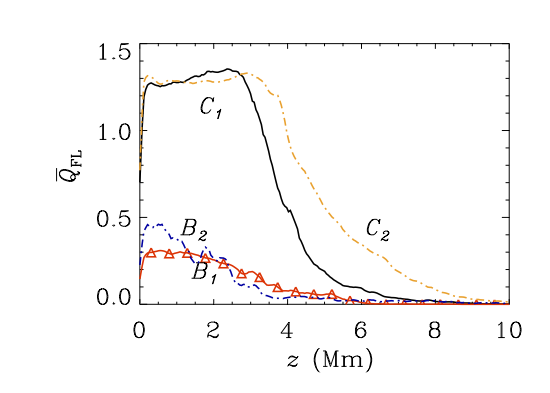}
\end{overpic}
\caption{\label{fig:qflbar} $\overline{Q}_{\mathrm{FL}}$ obtained by integrating the Ohmic heating, $Q_{\mathrm{FL}}$, over points inside the horizontal red boundary marked in Figs.~\ref{fig:qflB1C1} and ~\ref{fig:qflB2C2} for all the four flares.}
\end{figure}

\section{Applying the $WG_M$ method to simulated flare data} \label{Analyses}

\subsection{Implementation of $WG_M$ method} \label{WGMmethod}

We investigate the pre-flare behavior of the simulated 3D $\delta$-type sunspot by the tool put forward by \cite{Korsos2015}. Namely, they introduced the weighted horizontal magnetic gradient proxy (denoted as $WG_{M}$) between two opposite magnetic polarity umbrae in an $\delta$-spot, and demonstrated that $WG_{M}$ could be successfully applied to identifying pre-flare patterns above M5 energetic flare class. The $WG_{M}$ proxy is based on two components: (i) the total unsigned magnetic flux summed for all the considered umbrea of opposite polarities and (ii) the distance between area-weighted two barycenters of the positive and negative polarities within the entire $\delta$-spot. Initially, the $WG_{M}$ method was developed on a sample of 61 cases using the SOHO/MDI-Debrecen Data (also known as SDD) and further tested with the SDO/HMI-Debrecen Data (also known as HMIDD, the continuation of the SDD) catalogue in \cite{Korsos2016}.
 In empirical analyses, for all the observed flare cases, two flare pre-cursor patterns were discovered:

\begin{enumerate}
\item The pre-flare behavior of the $WG_{M}$ quantity itself exhibited characteristic patterns: increase, and the maximum value of the magnetic flux gradient followed by a gradual decrease prior to flaring. The aqua "inverted V-shape" points out the pre-flare behavior of the $WG_{M}$ in the top panels of Figs.~\ref{0level}-~\ref{3200level}.

\item The pre-flare behavior pattern of the distance parameter is based on the approaching-receding motion between the area-weighted barycentres of the positive and negative polarities  prior to flare. It was found that the evolution of distance has actually two ways to behave after the moment when the distance is regained its around value it had at the beginning of the approaching phase. One way is when the distance becomes decreasing and another way is when the distance keeps growing continuously before the flare occurrence. The duration of approaching-receding motion of the area-weighted barycenters of opposite polarities is highlighted by a red parabolic curve in the middle panels of Figs.~\ref{0level}-\ref{3200level}. A parabolic curve is fitted from the starting time of the approaching phase to the end of the receding phase, taking its minimum at the moment of reaching the closest position of the two barycenters derived from the minimum point of the data.
\end{enumerate}

In \cite{Korsos2015}, the next diagnostic tools were introduced to probe the pre-flare behavior patterns, where the viability of the diagnostic tools were tested on a sample of 61 cases observed during the SOHO/MDI era. 

\begin{itemize}

\item The first one is based on the relationship between the values of the maxima of the $WG_{M}$ ($WG^{max}_{M}$) and the highest GOES flare intensity class of ARs. In the case of the current, simulated artificial AR, presented here, the applicability of the intensity estimation should be made cautiously, because this relationship has yet been determined only for high-energy flares, i.e. above M5, while the simulated flares are B and C classes only. 

\item Next, the estimation of the flare onset time ($T_{est}$), is based on the relationship found between the duration of receding motion of the opposite polarities until the flare onset ($T_{D+F}$) and duration of the approaching motion ($T_{C}$) of the opposite polarities. K15 have also classified the selected spot groups of their study by age - into younger  or older than three days - and repeated the investigation separately for these two groups, in order to determine how fundamental this relationship may be. The following regression may be one of the most useful results found from the $WG_{M}$ method:

\begin{equation}
T_{est} = a \cdot T_{C} + b,  
 \label{time}
\end{equation}

where $a =1.29$ $(0.85)$ [hr] and  $b =1.11$ $(12.8)$ [hr] in the younger (older) than three-day case, respectively. Given that the eruptive  events in the simulation happen much faster, i.e. on the time scale of minutes rather than hours like in the real Sun due to the practical  limiting reasons on CPU access, we need to appropriately re-scale the hours to minutes time scale in Equation (\ref{time}). Furthermore, we also need to re-scale the three-day (= 72 hours) limit of what is labelled younger (older) emerging fluxes to appropriate minute-scale limit.  Now, given the linear structure of Equation (\ref{time}), we use $a =1.29$ $(0.85)$ [min] and $b =1.11$ $(12.8)$ [min] for the emerging fluxes younger (older) than 72 minutes.  

\item The last tool is the percentage difference ($WG^{\%}_{M}$) calculated between the values of pre-flare $WG^{max}_{M}$ and the values of $WG_{M}$ at the flare moment of onset ($WG^{flare}_{M}$). If $WG^{\%}_{M}$ may be over 54\%, no further flare of the same class or above would be expected; but, if $WG^{\%}_{M}$ is less than $\sim$42\%, further flares of the same class are probable within about an 18 hour window (i.e.,18 mins in the simulation).

\end{itemize}

For the pre-flare behaviour of the $WG_{M}$ and barycentric distance parameters to qualify as true pre-cursor event and not just fluctuation, it was introduced that the duration of i) the decrease in the distance parameter during the  approaching phase and ii) the increase in $WG_{M}$  has to take place for about at least a minimum of 4 hrs for real flares. These conditions were satisfied for over 90\% of the studied cases and were accepted as criteria for cut-off to remove fluctuations. Therefore, flux rising or approaching events with shorter durations were considered as fluctuations. Now, given the employed re-scaling for the simulation, the applied cut-off here is about 4 mins. iii) Further, for real solar applications the $\delta$-spot has to also satisfy to be within a certain belt around the central meridional, i.e. within $\pm$70 degree, however, this condition is not applicable here, given the chosen geometry providing a perpendicular view representing the solar surface. iv) Finally, and perhaps most importantly, for a pre-cursor event to qualify as a true pre-flare signature, it is required that (i) and (ii) must occur concurrently. I.e., based on the analysis of the 61 M5 or above flare cases, it was found by K15 that for all the studied samples (i) and (ii) were always present.

\subsection{Analysis and interpretation in terms of pre-flare dynamics} \label{Interpretation}

\begin{figure} [h!]
\centering
\plotone{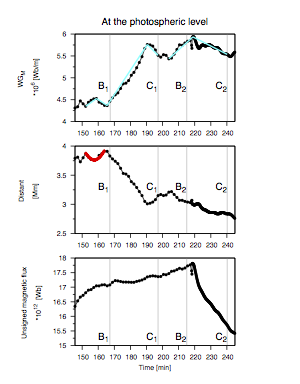} 
\caption{Evolution of various pre-flare indicators, applied to the simulation data. The {\it x}-axis is time [min]. (a) The upper panel shows the temporal variation of $WG_{M}$. The pre-flare behavior of the $WG_{M}$ is indicated by aqua "inverted V-shape", where a peak corresponds to a follow-up flare. (b) Middle panel demonstrates the evolution of the distance of the barycenters of opposite polarities. The red curve fit marks the full approaching-receding motion of the distance between area-weighted barycentres of opposite polarities. The vertical dashed lines indicate the moments when the flares occurred. Note, however, the flares do not occur at the photosphere (see e.g. Fig.\ref{fig:qflbar} to determine the height range for the flare location). (c) The bottom panel shows the evolution of the unsigned flux amounts.}
\label{0level}
\end{figure}

Let us now apply the $WG_{M}$ method to the numerically simulated flaring $\delta$-spot. We calculate the $WG_{M}$ in the entire $\delta$-spot like in the case of real sunspot data. The investigation in terms of the pre-flare dynamics starts from t=145.22 min, i.e. from the moment when the simulated AR finally emerged through the photosphere and developed into a complex set of loops.
From the simulation data we know, e.g. from constructing the temperature contour and $\overline{Q}_{\mathrm{FL}}$ plots at various heights, that all the flares occurred between 0.3--3.25 Mm in height (see Fig. \ref{fig:mag_temp}a--d and Fig. \ref{fig:qflB2C2} for $B_{1}$, $C_{1}$, $B_{2}$ and $C_{2}$ flares). These inspire us to extend and apply the flare pre-cursor identification analysis in the solar atmosphere {\it as function of height}, from the photosphere to as high as $z=3.6$ Mm. The aim is to demonstrate that the flare pre-cursor patterns may appear earlier in time, when applied to data available higher in the lower solar atmosphere, as compared to its counterpart form photospheric analysis.

\subsubsection{Investigation of pre-flare behavior at the different heights}

Let us now track the temporal variation of $WG_{M}$, distance of the area-weighted barycentres of the opposite polarities and the unsigned magnetic flux at the different heights in the lower solar atmosphere similar to the analysis carried out earlier with observed data at the photosphere, demonstrated in K15.

\underline{At the photosphere:}
From inspecting Fig.~\ref{0level}, we recognise the pre-flare patterns of $WG_{M}$ (aided by aqua "inverted V-shape" in Fig.~\ref{0level}) as follows: a rising phase, a first maximum value of the flux gradient (at 158.89 min, i.e a peak in the aqua line preceding the first flare) that is followed by a gradual decrease which culminates in the $B_{1}$ flare at $t=167.5$ min. About 8 mins later, after the first maximum value of the $WG_{M}$, one finds another (now a much more pronounced) steep rise and the associated high maximum value of the flux gradient (second aqua peak). This peak is followed, again, by a gradual decrease which ends with the $C_{1}$ energy flare. Another 10 mins later, from the $C_{1}$ flare, the $WG_{M}$ shows again a pre-flare behavior before the $C_{2}$ flare (i.e. third aqua peak). Unfortunately, in the case of the $B_{2}$ event (for ease and convenience marked as vertical dashed line) we cannot observe the complete pre-flare behavior of the $WG_{M}$. All can be said about it is that the $B_{2}$ flare happened during the rising phase of the $WG_{M}$ before the $C_{2}$ flare without pre-cursor signature in the data.

Let us now follow the evolution of the distance parameter in time in the data at photospheric level (middle panel of Fig. \ref{0level}). We can see the mark of approaching-receding motion of the area-weighted barycenters of opposite polarities  before the $B_{1}$ flare (indicated by the red parabola in Fig.~\ref{0level}). In the case of subsequent $C_{1}$, $B_{2}$ and $C_{2}$ flares, however, we cannot identify the complete pre-flare behaviors of the distance parameter using the simulation data available at photospheric level. For example, after reaching the minimum value during the approaching phase at $\sim$190 mins, the value of distance parameter did not increase enough during the receding phase to regain its (about the same value) at the start of the approaching, which is a prerequisite for applying the $WG_{M}$ method successfully.

We conclude, at this stage, that using the photospheric data, only the $B_{1}$ flare had the required concurrent qualifying pre-cursors for indicating the potential development of a flare. Although there are tempting pre-cursors for the $C_{1}$ flare, the distance parameter does show the required full parabolic U-shape.

\begin{figure} [h!]
\centering
\plotone{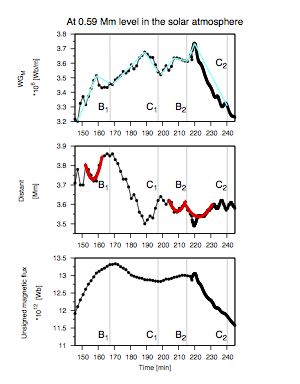} 
\caption{ The diagrams show the evolution of the same physical parameters for the artificial AR as of Fig. \ref{0level} but at the height of 0.59 Mm above the photosphere. }
\label{500level}
\end{figure}

\begin{figure} [h!]
\centering
\plotone{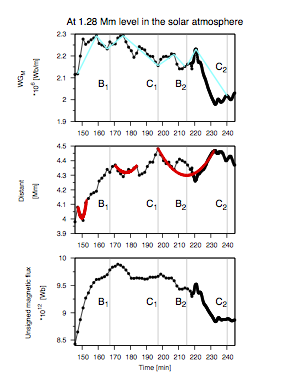} 
\caption{Same as Fig. \ref{0level} but at the height of 1.28 Mm above the photosphere. Here, the pre-flare evolution stages of $WG_{M}$ for the first two flares are not visible, there is indication only for the second C-class flare. }
\label{1000level}
\end{figure}

\underline{At 0.59 Mm level in the low chromosphere:}
 In Fig. \ref{500level}, we show the evolution of the three parameters ($WG_{M}$, distance and unsigned magnetic flux) in the low chromosphere. Further, signatures of first point to note is that: one more increasing and decreasing phase of $WG_{M}$ starts to appear before the $B_{2}$ flare, starting from $\sim$196 mins. The two additional approaching and receding phases of the distance parameter become identifiable, before the $B_{2}$ and $C_{2}$ flares, respectively. At this level of height, we found (though with some level of fluctuations present) the characteristic increasing and decreasing phase of $WG_{M}$ prior to each of these flares (see the aiding aqua lines for marling the four peaks). Also, we observe the signatures of the approaching-receding motion between the area-weighted barycenters of opposite polarities prior to $B_{1}$, $B_{2}$ and $C_{2}$ flares (marked with three red U-shapes). 
 
We conclude, at this stage using data at 0.59 Mm, that the pre-cursors became more pronounced for the $B_{1}$ flare; for $B_{2}$ we still cannot be fully certain that a flare may develop as the distance parameter does not satisfy the minimum 4 mins decrease criteria of U-shape. Although there are tempting pre-cursors for the $C_{1}$ flare, the distance parameter does show the required full parabolic U-shape.

\begin{figure} [h!]
\centering
\plotone{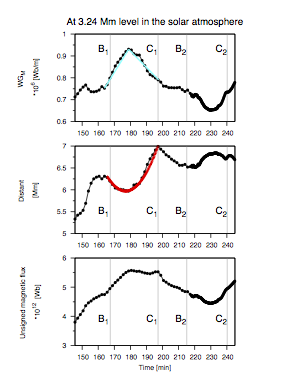} 
\caption{Same as Fig.~\ref{0level} but at 3.2 Mm high in the solar atmosphere. The two B-class flares are only marked for completeness, they cannot be confidently identified.}
\label{3200level}
\end{figure}

\underline{At 1.28 Mm in middle chromosphere:} When one ascends further up in the solar atmosphere and reaches the 1.28 Mm level, one sees changes in the evolution of the $WG_{M}$, distance between area-weighted polarity barycenters and the unsigned magnetic flux. It is found, at this height, that the pre-flare behaviour of $WG_{M}$ is difficult to recognize prior to  $B_{1}$, $C_{1}$, $B_{2}$ and $C_{2}$ flares but they are there and may qualify as pre-cursors.

In Fig.~\ref{1000level}, before the $C_{1}$ flare, the duration of the approaching-receding motion of the distance starts to form between 170 and 184 min but this interval will become longer in the higher solar atmosphere. The two approaching-receding phases of the distance identified at the 0.59 Mm level (for $B_{2}$, $C_{2}$) merge when ascending further to 1.28 Mm. It is also found, at this height, that the indicator of approaching-receding motion of the $B_{1}$ flare has actually started to disappear. The decrease is only 3.32 mins before the $B_{1}$ which does not satisfy the threshold criteria of minimum of 4 mins decrease.

\underline{At 3.24 Mm above the photosphere:} As one ascends even higher, one finds that the evolution of  $WG_M$ and distance changes remarkably (see Fig. \ref{3200level}) when compared to their behaviour at the photosphere (Fig. \ref{0level}). Here, we also note that the pre-flare behaviour of $WG_{M}$ is recognisable between 145.56 and 161 mins which could link to $B_{1}$ but we avoid the analysis of B-class flares at this level based on the plateaus of the $\overline{Q}_\mathrm{FL}$ during flares shown in the simulations (see Fig.~\ref{fig:qflbar}). We also cannot recognize anymore any meaningful characteristic pre-flare behaviors of the distance prior to these two small flares.

In the two C flare cases, when the transition region and the lower corona is reached at this height, we do recognise, however the following properties of the $WG_M$ and the distance between the area-weighted polarity barycenters: (i) First of all, the steep rise from 164 mins and a high maximum value of the  weighted horizontal gradient of the magnetic field is still followed by a less steep decrease prior to $C_{1}$ flare (see the aqua, "inverted V-shape"). The $WG_M$ has only rising phase before the $C_{2}$ flare at this height. (ii) The approaching and the receding characteristic features of the distance prior to $C_{1}$ flare are also there, but the distance parameter does not comply to be a qualifying criteria before the $C_{2}$ flare.

Based on the analysis of data available at the very high end of the lower solar atmosphere (i.e. at 3.24 Mm), we conclude that pre-flare signatures of $C_{1}$ can be finally confirmed (as opposed to the cases at lower heights discussed earlier). Signatures of the small B-flares are not clear and neither are they for the $C_{2}$ flare. 

Finally, similar to observed data of real sunspots, the unsigned magnetic flux (lower panel Figs.~\ref{0level}-\ref{3200level}) does not show any special behavior to be useful for flare pre-cursor. 

\subsubsection{Optimum height(s) search for an earlier flare pre-cursor identification}

\begin{table*} [ht!]
\caption {Summary table of the investigated properties of the two B- and two C-class flares at their optimum heights.}
\centering
\begin{tabular}{cccccccccc}

\hline
\hline
Flare &Interval&Optimum height &$\overline{Q}_\mathrm{FL}$&$WG_{M}^{Max}$  &$WG^{flare}_{M}$  & $T_{C}$  &$T_{D+F}$   &   $T_{est}$ & $WG^{\%}_{M}$\\
&&[Mm]&[Mm]& $\cdot$$10^{6}$ [Wb/m]  &$\cdot$$10^{6}$[Wb/m] & [min] & [min] &   [min] & [\%]\\
\hline
\hline

			  \multirow{2}{*}{$B_{1}$ }&Min   &0.3    & 0.1  &  3.99    	&  3.93	& 3.34 	& 11.94	 & 5.40	& 1.5\%	 \\
                            &Max  &  0.4   &  2  &  3.87    	&  3.80	& 3.34 	& 11.94 	 & 5.40	& 2\%	 \\
\hline

			  \multirow{2}{*}{$C_{1}$} &Min   &2.3   &  0.1  &  1.36    	&  	1.18	& 11.60 	&  19.98	 & 	16.20	&13.2\%	 \\
                            &Max  &2.9     &  3 &  1.05	   	&  	0.89	& 11.60 	&  19.98	 & 	16.20		&15.2\%	 \\
\hline

			  \multirow{2}{*}{$B_{2}$ }&  Min & \multirow{2}{*}{0.5} &  \multirow{2}{*}{0.5}  &   \multirow{2}{*}{3.92}	&  	  \multirow{2}{*}{3.92}   	&   \multirow{2}{*}{3.34}	 &    \multirow{2}{*}{10}  &	  \multirow{2}{*}{14.7}	& \multirow{2}{*}{0.1\%}	 \\
			             &  Max &                                &    &   	                  	&  	&  &   &	&	 \\
\hline	

			  \multirow{2}{*}{$C_{2}$} &Min   &1.2  &  0.1   &  2.23  	&  	2.01	&15.70 & 	25.70  &25.40	&9.5\%	 \\
                            &Max  &1.8  &  4  &  1.54	   	&  	1.39	&16.70 & 	24.70  &26.2		&10\%	 \\
\hline

\hline
\end{tabular}

\label{table3}
\end{table*}

The evolution of the $WG_M$ and the distance of the area-weighted barycentres of opposite polarities are different at various heights, as has been described above. In order to improve the flare pre-cursor capability of the $WG_{M}$ method therefore we try to identify optimum height(s) in the solar atmosphere. The investigated heights are where the pre-cursor behaviours of the $WG_{M}$ and distance parameters are identifiable prior to each flare. The optimum height(s) would be where the distance parameter would yield the earliest sign of pre-flare behavior in time. Table \ref{table3} summarises the key parameters for finding the optimum heights.

First, in Fig.~\ref{height} we plot the variation of the start time of the approaching phase (green lines), the moment of the closest approach (blue lines) and the estimated flare onset time (magenta lines) as function of height. In Fig.~\ref{height}, the filled square/triangle/circle/star symbols mark the calculated corresponding data of $B_{1}$/$C_{1}$/$B_{2}$/$C_{2}$-class flare. The black vertical lines indicate the onset time of the flares, where the strength of the flare ($B_{1}$/$C_{1}$/$B_{2}$/$C_{2}$) is labelled on the top axis. The grey strips mark the vertical extent where ohmic heating  of the "current carrying" field lines reach plateaus of the $\overline{Q}_\mathrm{FL}$ during flares in the simulations (see Fig.~\ref{fig:qflbar}). Most noticeable is that, in general, there are certain heights above the photosphere, where the approaching motions begin earlier and reach the closest point of approach also earlier than at the photosphere or at other heights in the solar atmosphere.

 In Fig.~\ref{height}, the start time of the approaching phase (first green line with squares) of the $B_{1}$ flare is sooner and it also reaches the moment of the closest approach sooner (first blue line with squares) between heights at 0.3-0.4 Mm than at the photosphere or at any other heights. In the case of the $B_{2}$ flare, the optimum height, i.e. having the earliest time of beginning of approach, seems to be 0.5 Mm. Similarly, for the $C_{1}$ flare the start time of the approaching phase and moment of the closest approach is earliest between heights 2.3 and 2.9 Mm above the photosphere. We can clearly see that the start time of the approaching phase and the moment of closest approach corresponding to the $C_{2}$ flare is earliest between heights 1.2 and 1.8 Mm from the photosphere. This result is rather important: if we are able to identify the optimum height where the moment of start time of the approaching phase as well as the moment of closest approach is indeed earlier than at any other heights in the solar atmosphere, then the analysis carried out at this height may (hopefully considerably in practice) improve the capacity of flare pre-cursor capability, e.g. yielding a more accurate flare onset time. Furthermore, it also seems that the optimum height may depend on the energetic flare class. This could be a significant progress if confirmed by  observations on a larger database. 

\begin{figure} [h!]
\centering
\epsscale{1.2}
\plotone{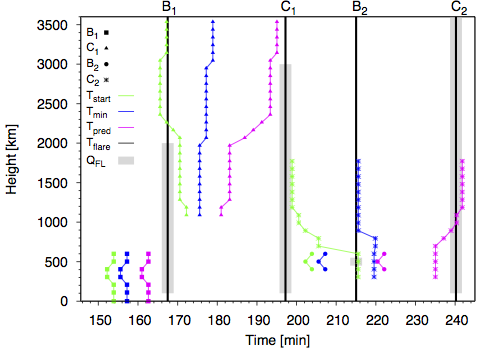} 
\caption{The filled square/triangle/circle/stars symbols are associated with $B_{1}$/$C_{1}$/$B_{2}$/$C_{2}$-class flares, respectively. The actual moment of start times of approaching (green lines), times of momentum of the closest approaching point between two barycenters (blue lines) and the estimated flare onset time by Equation (\ref{time}) (magenta lines) are plotted as function of height. The black vertical lines denote the two B-class and two C-class flares (at 167.5, 215.03, 197.2 and 240.2 min). The grey areas demonstrate the height extent where the ohmic heating of their "current carrying" field lines reach more than 95\% of the maximum ($Q_{FL}$) near the onset time of the two B-class and the two C-class flares, respectively. }
\label{height}
\end{figure}

In Table \ref{table3}, we list some properties of the flares determined at the minimum and maximum value of their optimum heights and the minimum and maximum height values corresponding the plateaus of the $\overline{Q}_\mathrm{FL}$. Table \ref{table3} includes the maximum value of the $WG_{M}$, value of $WG_{M}$ at the flare onset, duration of the simulated compressing phase ($T_{C}$) and receding motion until flare onset ($T_{D+F}$), the estimated flare onset time ($T_{est}$) elapsed from the moment of reaching the closest point during the approaching-receding motion to the flare (computed from Eq.~\ref{time}) and the ratio of maximum value of the $WG_{M}$ to the value of the $WG_{M}$ at flare onset. The estimated onset time and the elapsed time of simulated receding motion until flare onset are close to each other for the two B- and C-class flares at the optimum heights.

First, if we apply $T_{C}$ obtained from the first approaching-receding motion of the barycentric distances before the $B_{1}$ flare between heights of 0.3 and 0.4 Mm then the time difference is 6.54 min between the estimated and simulated flare onset time (see the values in Table \ref{table3}). The $C_{1}$ flare occurred only 3.78 min later than the expected onset time. For the $B_{2}$ flare the time difference between the estimated and the observed onset time is a mere 4.7 minutes. The onset time of $C_{2}$ flare is estimated well because it occurred only 1.5 min later than one may expect it from mere photospheric data. 
Also, the minimum and maximum values of the optimum heights of $B_{1}$, $C_{1}$, $B_{2}$ and $C_{2}$ flares are in the plateau ranges of the $\overline{Q}_\mathrm{FL}$ (see the values in Table \ref{table3}).

Last, we investigate the percentage difference ($WG^{\%}_{M}$) at identified optimum heights. The $WG^{\%}_{M}$ does not seem to be applicable to the simulation data, unlike to
observational data. The values of $WG^{\%}_{M}$ are small which means that one may expect further flare(s) during the decreasing phase of $WG_{M}$, but this is not taking place. So, further investigation may be needed to exploit the applicability of this parameter.

\section {Conclusion} \label{conclusion}

\cite{Piyali2016} modelled a $\delta$-sunspot like feature formed due to the collision of two magnetic regions with opposite polarity arising from the interaction of emerging magnetic flux with stratified convection. The two opposite polarities of the magnetic field are part of the same initial subsurface structure and their collision causes recurring flaring.

 Four flares were simulated, classified as two B- ($B_{1}$ and $B_{2}$) and two C- ($C_{1}$ and $C_{2}$) classes. To this flaring simulated AR, we have applied the $WG_{M}$ method, put forward by \cite{Korsos2015} in the context of identifying flare pre-cursors, tracked the temporal evolution of the $WG_{M}$, the variation of distance between the barycenters of opposite polarities and the unsigned magnetic flux at different heights in the model solar atmosphere from photosphere up to 3.6 Mm. We identified two important pre-flare behaviors, at stepping intervals of 100 km in height in the solar atmosphere: (i) Foremost, the typical and characteristic pre-flare variation of $WG_{M}$ was confirmed and found to begin with an increase of its value until a maximum, followed by a decrease until the flare(s) onset. The pre-flare behavior of $WG_{M}$ was found to be height-dependent; at some heights they were vague and less easy (or impossible) to identify, while at another heights the behavior was clearly identifiable. The height-variation of the clarity of the pre-flare behavior of $WG_{M}$ was also found to be dependent on the strength of the simulated flare. The B-class energetic flares showed a stronger clarity in terms of the pre-flare behavior at lower solar atmospheric heights when compared to their counterparts of C-class eruptions. An important common property was also found, namely, the eruption occurred on the decreasing phase of $WG_{M}$ in all cases. Therefore, the simulation and the application of the $WG_{M}$ method to the modelled AR is consistent with the preliminary results found when applying this method to real solar data of stronger than M5 flares \citep{Korsos2015}. This finding encourages to suggest that the observational detection of such behavior in $WG_{M}$ may serve as a useful and practically simple alert tool for eruption(s) about to occur. (ii) Secondly, the approaching-receding motion, i.e., the decreasing-increasing distance between the area-weighted polarity barycenters prior to flare(s) seems to be another applicable indicator of an impending flare. Similarly to the pre-flare behavior of the $WG_{M}$, the approaching-receding motion was height-dependent prior to flares; at some heights they were vague and less easy to identify, while at another heights the behavior was clearly identifiable. In general, one may state that the clearest identification was at lower solar atmospheric heights for the B-class flares while for the C-class flares it was higher up in the chromosphere.
 
Next, we investigated the variation of the moment of start time of the approaching phase, the moment of the closest approach and estimated flare onset time as a function of height (see the Fig \ref{height}). This investigation was carried out by searching for specific heights at which the approaching motion of the area-weighted barycentres of the opposite polarities corresponding to a flare event would start earlier and reach its closest approach distance earlier than at any other level (well, at least photospheric level) in the solar atmosphere, so that we may estimate the onset earlier in time. Also, the pre-flare behavior of $WG_{M}$  can be recognized at the optimum heights.

In Fig. \ref{height} and from Table \ref{table3}, the start time of the approaching phase and time of closest approach of the area-weighted barycentres of opposite polarities corresponding to the $B_{1}$ flare takes place earliest between heights 0.3-0.4 Mm than at any other (especially  photospheric) level. The estimated onset time ($T_{est}$) of the $B_{1}$ flare is 5.4 min, according to Eq.~(\ref{time}), between 0.3 and 0.4 Mm compared to the simulated $T_{D+F}$ of 11.94 min. In case of the $B_{2}$ flare, the optimal height is about 0.5 Mm (i.e. low chromosphere) where the difference is 4.7 minutes between the simulated occurrence time and the estimated onset time. For the $C_{1}$ flare the moment of start time of the approaching phase and moment of reaching the minimum point happened earliest between heights of 2.3 and 2.9 Mm when compared to their photospheric counterparts. From Table \ref{table3}, we can see that the $C_{1}$ flare occurred 3.78 min later than the estimated onset time. Furthermore, in Fig.~\ref{height}, we can also see that the moment of start time of the approaching phase of the $C_{2}$ flare starts earlier, and reached the closest approach point earlier, at 1.2 Mm measured from the photosphere. Here, the onset time for the $C_{2}$ flare is well estimated because the flare took actually place only 1.5 min later than estimated (see Table \ref{table3}). The shaded grey areas shown in Fig.~\ref{height} correspond to the flare initiation height estimates, made using  the full width at $95\%$ of the maximum of the Ohmic heating curve as defined by $Q_\mathrm{FL}$, as a function of height. Hence, when comparing the last three rows of Table~\ref{tab:flareonset}, we find that for all but the $B_{1}$ flares, the estimates from all three methods - temperature anomaly (see Fig.~\ref{fig:mag_temp}a-d), Ohmic heating peaks (see Fig.~\ref{fig:qflbar}), $WG_M$ - agree well with each other. For flare $B_1$, the temperature anomaly estimate does not agree with the estimates from the Ohmic heating peak location and the $WG_M$ method.  
We suspect that there may exist a relation between the optimum height for earliest estimation of the flare by the $WG_M$ method and the flare initiation height from the analysis of temperature and Ohmic heating signatures. 

In brief summary, we found that the typical pre-flare dynamics reported in K15 does seem to work for the simulated low energy flare events, as seen in this case study mimicking the evolution of an AR. Our initial results are encouraging because we do observe very similar pre-flare behavior of the $WG_{M}$ and the distance parameter between the polarity barycenters in real sunspot data K15 as well, indicating that the predictive temporal behavior of these parameters may indeed be an intrinsic feature of the physical processes preceding flare onset. The fact, that the application of the $WG_M$ method developed by K15 gives similar results for observed (GOES C-, M- and X-class) flares as well as the simulated (B- and C-class) flares also gives us confidence that a basic physical mechanism of flare initiation has been {\it phenomenologically} captured reasonably well in the flare simulation reported in \cite{Piyali2016}. The other interesting aspect worth to mention is that the flare pre-cursors are height and flare strength dependent. Unfortunately, we cannot give a proper physical explanation for this behaviour yet as this would require a more in-depth study of the reconnection process itself  that is beyond the scope of the current paper. The height-dependent behaviour of the flare pre-cursors may be linked to the so-called push and pull reconnection, observed in laboratory plasma experiments \citep{Yamada2010}. For a more definite and conclusive statement one may need to carry out an ensemble of simulations of the evolution of $\delta$-sunspots with flares of higher GOES class (M-, and X-classes) and test this relation as well as flare pre-cursor capability of the $WG_M$ method on a statistically significant sample of simulated and observed sunspots. 

  \section{ Acknowledgements} 
     We thank the anonymous referee for a very careful reading of our manuscript which has led to a marked improvement in the clarity of this paper. MBK is grateful to the University of Sheffield and the Hungarian Academy of Sciences for the support received. MBK also acknowledges the open research program of CAS Key Laboratory of Solar Activity, National Astronomical Observatories, No. KLSA201610. PC thanks the University of Sheffield for hospitality and support for a visit during which this work was initiated and also the CAS PIFI project 2017VMC0002 and National Astronomical Observatories, Beijing for support. The simulation was carried out on NASA's Pleiades supercomputer under GID s1061. PC also acknowledges computing time awarded on PARAM Yuva-II supercomputer at C-DAC, India under the grant name Hydromagnetic-Turbulence-PR.  We have used the 3D visualisation software Paraview for volume rendering and field line plotting. RE is grateful to Science and Technology Facilities Council (STFC, grant nrs ST/L006316/1 and ST/M000826/1) UK and the Royal Society for their support. The authors also express their gratitude towards Christopher J. Nelson and Michael S. Ruderman  (both at University of Sheffield, U.K.) for a number of useful discussions and improving the manuscript.


\begin{thebibliography}{}


\bibitem[\protect\citeauthoryear{Archontis \& Hood}{2008}]{Archontis2008} Archontis, V. \& Hood, A. W. 2008, \apj, 674, L113

\bibitem[\protect\citeauthoryear{Aschwanden}{2005}]{Aschwanden2005} Aschwanden, M. J., Physics of the Solar Corona. An Introduction with Problems and Solutions (2nd edition), 2005, ISBN: 3-540-30765-6


\bibitem[\protect\citeauthoryear{Burtseva \& Petrie}{2013}]{Burtseva2013} Burtseva, O., Petrie, G., 2013, \solphys, 283, 2, 429

\bibitem[\protect\citeauthoryear{Chatterjee {\it et al.}}{2016}]{Piyali2016}  Chatterjee, P., Hansteen, V. \&  Carlsson, M., 2016,  Physical Review Letters, 116, 10, 101101

\bibitem[\protect\citeauthoryear{Cheung \& DeRosa}{2012}]{Cheung_DeRosa2012} Cheung, M. C. M. \& DeRosa, M. L. 2012, \apj, 757, 147

\bibitem[\protect\citeauthoryear{DeVore \& Antiochos}{2008}]{DeVore2008} DeVore, C. R. \& Antiochos, S. K., 2008, \apj, 680, 740

\bibitem[\protect\citeauthoryear{Evershed}{1910}]{Evershed1910} Evershed, J., 1910, MNRAS, 70, 217

\bibitem[\protect\citeauthoryear{Falconer {\it et al.}}{2008}]{Falcon2008} Falconer, D. A., Moore, R. L., \& Gary, G. A.  2008,  \apj, 689, 1433

\bibitem[\protect\citeauthoryear{Green {\it et al.}}{2011}]{Green2011} Green, L.M., Kliem, B., Wallace, A.J., 2011, Astron. Astrophys., 526, A2.

\bibitem[\protect\citeauthoryear{Isobe {\it et~al.}}{2005}]{Isobe_etal2005} Isobe, H., Takasaki, H. \&  Shibata, K. \apj, 2005, 632, 1184

\bibitem[\protect\citeauthoryear{Kempf}{1910}]{Kempf1910} Kempf, P., 1910, Astron. Nachr., 185, 197

\bibitem[\protect\citeauthoryear{Kors\'os {\it et al.}}{2015}]{Korsos2015} Kors\'os, M. B., Ludm\'any, A., Erd\'elyi, R., Baranyi, T., 2015, \apj, 802, L21

\bibitem[\protect\citeauthoryear{Kors\'os \& Ruderman}{2016}]{Korsos2016} Kors\'os, M. B. \& M. S., Ruderman, 2016, Ground-based Solar Observations in the Space Instrumentation Era ASP Conference Series, 504, 43-47

\bibitem[\protect\citeauthoryear{K\"unzel}{1960}]{Kunzel1960} K\"unzel, H., 1960, Astron. Nachr., 285, 271

\bibitem[\protect\citeauthoryear{Leka {\it et al.}}{1996}]{Leka1996} Leka, K. D., Canfield, R. C., McClymont, A. N., van Driel-Gesztelyi, L., 1996, \apj, 462, 547

\bibitem[\protect\citeauthoryear{Li {\it et al.}}{2005}]{Li2005} Li, J., Mickey, D. L.,  LaBonte, B. J., 2005,  \apj, 620, 1092

\bibitem[\protect\citeauthoryear{Livi {\it et al.}}{1989}]{Livi1989}Livi, S.H.B., Martin, S.,Wang, H., Ai, G., 1989, \solphys, 121, 197

\bibitem[\protect\citeauthoryear{MacTaggart \& Hood}{2009}]{MacTaggart2009} MacTaggart, D. \& Hood, A. W., 2009, A\&A, 508, 445

\bibitem[\protect\citeauthoryear{Manchester \& Low}{2000}]{Manchester2000} Manchester, W., IV, \& Low, B. C., 2000, Phys. Plasmas, 7, 1263

\bibitem[\protect\citeauthoryear{Manchester}{2004}]{Manchester2004} Manchester, W., IV, Gombosi, T., DeZeeuw, D., Fan, Y., 2004, \apj, 610, 588

\bibitem[\protect\citeauthoryear{Takizawa \& Kitai}{2015}]{Takizawa2015}Takizawa, K. \& Kitai, R., 2015, \solphys, 290, 7, 2093

\bibitem[\protect\citeauthoryear{Tanaka}{1975}]{Tanaka1975} Tanaka, K., 1975, BBSO Preprint No.0152, Big Bear Solar Observatory

\bibitem[\protect\citeauthoryear{Sammis \& Zirin}{2000}]{Sammis2000}Sammis, I., Tang, F., Zirin, H.,  2000, \apj, 540, 583.

\bibitem[\protect\citeauthoryear{Savcheva {\it et al.}}{2012}]{Savcheva2012} Savcheva, A. S., Green, L. M., van Ballegooijen, A. A., DeLuca, E. E., 2012, \apj, 759, 2, 105

\bibitem[\protect\citeauthoryear{Schrijver}{2007}]{Schrijver2007} Schrijver, C.,  2007, \apj,  655, L117	

\bibitem[\protect\citeauthoryear{Selwa {\it et al.}}{2012}]{Selwa2012}Selwa, M., Poedts, S., DeVore, C. R., 2012, \apj, 747, 2, L21

\bibitem[\protect\citeauthoryear{Shibata \& Magara}{2011}]{Shibata2011} Shibata, K. \& Magara, T., 2011, Living Reviews in Solar Physics, 8, 6

\bibitem[\protect\citeauthoryear{Sterling {\it et al.}}{2010}]{Sterling_etal2010} Sterling, A. C., Chifor, C., Mason, H. E., Moore, R. L. \&  Young, P. R., 2010, A \& A, 521, 49

\bibitem[\protect\citeauthoryear{van Driel-Gesztelyi {\it et al.}}{1997}]{Lidia1997} van Driel-Gesztelyi, L., Csepura, G., Schmieder, B., Malherbe, J.-M., Metcalf, T., 1997, \solphys, 172, 151

\bibitem[\protect\citeauthoryear{Wang \& Shi}{1993}]{Wang1993} Wang, J. \& Shi, Z., 1993, \solphys, 143, 119

\bibitem[\protect\citeauthoryear{Yamada {\it et al.}}{2010}]{Yamada2010} Yamada, M., Kulsrud, R., Ji, H.  2010, Rev. Mod. Phy., 82, 603

\bibitem[\protect\citeauthoryear{Yan {\it et al.}}{2008}]{Yan2008}Yan, X. L., Qu, Z. Q., Kong, D. F., 2008, MNRAS, 391, 1887



\end{thebibliography}
\end{document}